\documentclass[conference]{IEEEtran}

\usepackage{cite}
\usepackage{graphicx}
\graphicspath{ {img/} }

\usepackage{algorithm}
\usepackage[noend]{algpseudocode}
\usepackage{bbm} 
\usepackage{subcaption}
\usepackage[shortlabels]{enumitem}
\usepackage{xcolor}
\usepackage{url}
\usepackage{amsfonts}
\usepackage{amsmath}
\usepackage{booktabs}
\usepackage{balance}

\begin{document}

\title{Driving with Data in the Motor City: Understanding and Predicting Fleet Maintenance Patterns}

\author{\IEEEauthorblockN{Josh Gardner\IEEEauthorrefmark{1}, Jawad Mroueh\IEEEauthorrefmark{2}, Natalia Jenuwine\IEEEauthorrefmark{2}, Noah Waverdyck\IEEEauthorrefmark{2},} 
\IEEEauthorblockN{Samuel Krassenstein\IEEEauthorrefmark{3},  Arya Farahi\IEEEauthorrefmark{2}, Danai Koutra\IEEEauthorrefmark{2}} \IEEEauthorblockA{\IEEEauthorrefmark{1}University of Washington; jpgard@cs.washington.edu}
\IEEEauthorblockA{\IEEEauthorrefmark{2}University of Michigan;
 [jmroueh, najenu, nweaverd, aryaf, dkoutra]@umich.edu} 
 \IEEEauthorblockA{\IEEEauthorrefmark{3}City of Detroit - Operations and Infrastructure Group}}

\maketitle

\begin{abstract}

The City of Detroit maintains an active fleet of over 2500 vehicles, spending an annual average of over \$5 million on purchases and over \$7.7 million on maintenance. Modeling patterns and trends in this data is of particular importance to a variety of stakeholders, particularly as Detroit emerges from Chapter 9 bankruptcy, but the structure in such data is complex, and the city lacks dedicated resources for in-depth analysis. The City of Detroit's Operations and Infrastructure Group and the University of Michigan initiated a collaboration which seeks to address this unmet need by analyzing data from the City of Detroit's vehicle fleet. This work presents a case study and provides the first data-driven benchmark, demonstrating a suite of methods to aid in data understanding and prediction for large vehicle maintenance datasets. We present analyses to address three key questions raised by the stakeholders, related to discovering multivariate maintenance patterns over time; predicting maintenance; and predicting vehicle- and fleet-level costs. We present a novel algorithm, PRISM, for automating multivariate sequential data analyses using tensor decomposition. This work is a first of its kind that presents both methodologies and insights to guide future civic data research.\footnote{For reproducibility details, including hyperparameter settings, software, and links to complete PARAFAC results, see the arXiv version of this paper.}
\end{abstract}

\section{Introduction}

On July 18, 2013, the City of Detroit (hereafter, simply Detroit) filed for Chapter 9 bankruptcy and initiated a recovery plan. The recovery plan includes major investments to update the police, fire, and emergency medical services departments and their fleets. Under this plan, the city is investing approximately 
\$447M over the next 10 years for the replacement and modernization of vehicle fleets and facilities.\footnote{\url{http://www.government-fleet.com/channel/procurement/news/story/2014/02/detroit-bankruptcy-plan-calls-for-fleet-modernization.aspx}}  Detroit manages and maintains a fleet consisting of over 2500 active vehicles, with four shops, six fuel sites, and 70 technicians to maintain the fleet. These vehicles are particularly critical to service delivery in the city, which has its population of over 672,000 spread over 139 square miles---an area larger than the City of Philadelphia with less than half of the population density.\footnote{\url{https://www.metrotimes.com/media/pdf/detroit_future_city_-_139_square_miles.pdf}}

Detroit spent an annual average of \$7.7M on maintenance and over \$5M on new vehicle purchases between 2010 and 2017.\footnote{These figures are based on the data used in this work.} Historical maintenance and purchase data can be utilized to efficiently allocate resources during the recovery effort.  
However, Detroit, like most municipalities, struggles with insufficient financial resources and capacity to analyze historical data and provide data-driven insights for decision-makers. 

\begin{figure}[t!]
    \centering
    \includegraphics[width = \columnwidth]{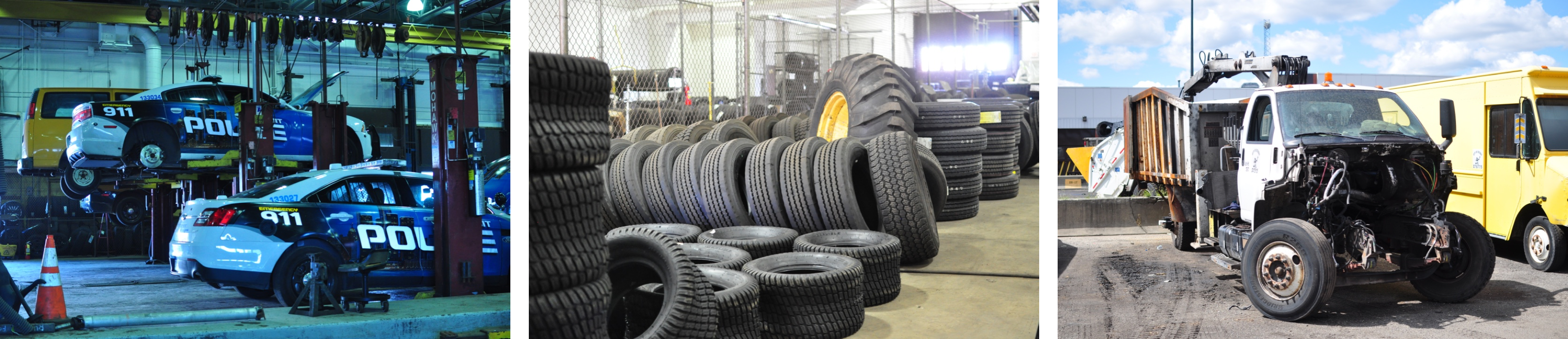}
    \caption{Vehicle fleet maintenance in Detroit.}
    \label{vehicle-fig}
\end{figure}

To fill this gap, the University of Michigan partnered with Detroit's Operations and Infrastructure Group. This collaboration has the dual goal of providing methods for data \textit{understanding} and \textit{prediction}, driven by three key research questions: {\bf(RQ1)} How can we uncover, validate, and interpret complex, multivariate patterns from fleet maintenance records? {\bf(RQ2)} Can we predict required vehicle maintenance? {\bf(RQ3)} Can we predict vehicle- and fleet-level maintenance costs?  

Answering these questions provides methods and interpretable algorithmic insights which will allow the city to better navigate the complex logistical and financial decisions all municipal governments face, including: optimize the allocation of existing resources; improve service delivery; reduce costs, fraud, and erroneous data; and make informed decisions about maintenance scheduling and future investments. For instance, when a vehicle is being repaired, it is unavailable for use, and is a stranded asset that reduces the city's capacity to deliver services. To ensure that the necessary types of vehicles are available when needed, the city must always maintain a surplus of vehicles, which result in added cost. The analyses in this work can address these issues: a multivariate analysis identifies common system repair patterns over time which assists technicians and analysts in understanding the fleet, informs technician hiring and allocation, and guides future vehicle deployment and procurement decisions; a predictive maintenance model proactively identifies necessary maintenance and can be used to optimize vehicle downtime, fleet availability, and job allocation across technicians and garages; and finally a cost forecasting model informs budgeting, resource allocation, and investment decisions.  

We address our research questions by developing and applying algorithms for multidimensional pattern extraction. Our main contributions are summarized as follows:
\begin{itemize}
    \item \textbf{Novel Study:} Vehicle maintenance data has not been evaluated in prior published data mining research. Our study sets a precedent for future research in this domain and provides the first data-driven approach.
    \item \textbf{Descriptive Analysis:} We use tensor decomposition and differential sequence mining, including the novel PRISM algorithm which presents a unified Bayesian approach these tasks, to discover complex vehicle-system-time repair patterns and their characteristic subsequences (\S~\ref{sec:parafac}). PRISM is the first algorithm to explicitly leverage the sequential nature of data modeled using the parallel factors decomposition (PARAFAC).
    \item \textbf{Predictive Analysis:} We leverage sequence neural networks to predict police vehicle maintenance and perform time series modeling to forecast vehicle- and fleet-level cost (\S~\ref{sec:lstm}). 
    \item \textbf{Guidelines \& Reproducibility:} We describe the challenges of data and analysis in real-world public-sector contexts and conclude with the lessons learned from our partnership (\S~\ref{sec:conclusions}). While a non-disclosure agreement with the City of Detroit prevents us from making the data publicly available, we release our code publicly so other municipalities and researchers can reproduce this work with their own data: \url{https://github.com/jpgard/driving-with-data-detroit/}. 
\end{itemize}

\section{Dataset}
\label{sec:data}

We analyze a comprehensive dataset of the entire Detroit-owned vehicle fleet and their maintenance jobs, provided by the Operations and Infrastructure Group in the City of Detroit. 
The records contain a mix of data transferred from prior paper records (with the oldest vehicle records dating to 1944) and those entered by new electronic record-keeping systems. Data entry is performed by several stakeholders, including maintenance technicians, managers, and analysts. 
The data consists of two tabular data sources.

The \textbf{vehicles table} (Table \ref{vehicles-table-supplementary}) 
consists of records, one per vehicle, representing every known vehicle currently or previously owned by Detroit. The table has information about each vehicle's manufacture, purchase, and use. It tracks data for police cars, garbage trucks, freight trucks, ambulances, boats, motorcycles, mowers, and other vehicles. The \textbf{maintenance table} (Table~\ref{maintenance-table-supplementary}) 
consists of job-level records for every individual maintenance job performed on any vehicles owned by Detroit. It includes everything from routine inspections, tire changes, and preventive maintenance to major collision repairs, glass work, and engine replacements.

\begin{table}[h!] 
\centering
\caption{Description of the \textit{vehicles} table.}
\label{vehicles-table-supplementary}
\resizebox{\columnwidth}{!}{
\begin{tabular}{ p{2.1cm} p{4.1cm} p{1.3cm}}
\toprule
{\bf Field} & {\bf Description} & {\bf Example} \\ \midrule
Unit\# & Unique Vehicle Identifier & 026603 \\
Dept\# & Code of dept vehicle is assigned to & 37 \\
Dept Desc & Description of department & POLICE \\
Make & Vehicle make & CHEVROLET \\
Model & Vehicle model & 2500 \\
Year & Model year of vehicle & 2002 \\
Last Meter & Odometer reading at last check (mi) & 52738 \\
Last Fuel Date & Most recent refuel 
& 2009-11-05 15:37:25 \\
Purchase Cost & Purchase cost, in US \$ & \$20,456 \\
Status Code & A = Active; S = Disposed & A \\
Status Desc & Description of status & Active Unit \\
LTD Maint. Cost & Total maintenance cost to date, in US \$  & \$5,951.04 \\
LTD Fuel Cost & Total fuel cost to date, in US \$  & \$9,295.01 \\
LTD Fuel Gallons & Total fuel consumption to date & \$3,646.6 \\ \bottomrule
\end{tabular}
}
\end{table}

\begin{table}[t!] 
\centering
\caption{Description of the \textit{maintenance} table.}
\label{maintenance-table-supplementary}
\resizebox{\columnwidth}{!}{
\begin{tabular}{p{2.2cm} p{3.7cm} p{1.5cm}}
\toprule
{\bf Field} & {\bf Description} & {\bf Example} \\ \midrule
Job ID & Unique identifier for job & 847956 \\
Year Completed & Year of completion & 2017 \\
Unit No & Vehicle identifier & 067602 \\
Work Order No & Unique identifier for work order & 635864 \\
Open Date & Work Order Open & 2017-01-17 \\
Completed Date & Work Order Completion & 2017-01-17 \\
Work Order Loc. & Location of work order & CODRF \\
Job Open Date & Job Open & 2017-01-17 \\
Job Reason & Job reason code & B \\
Job Reason Desc & Job reason description & BREAKDOWN/ REPAIR \\
Completed Date & Date Job Completed & 2017-01-17 \\
Job Code & Job ID & 24-13-000 \\
Job Description & Detailed description of job & REPAIR Brakes \\
Labor Hours & Hours of labor completed on job & 6.35 \\
Actual Labor Cost & Total cost of labor for job & \$348.16 \\
Commercial Cost & Commercial (non-city) labor & \$0 \\
Part Cost & Cost of parts for job & \$57.55 \\
Primary Meter & Odometer at repair time (mi) & 48250 \\
Job Status & Status code; DON = Done & DON \\
Job WAC & Job type code & 24 \\
WACDescription & Job type description & REPAIR \\
Job System & Code for vehicle system repaired & 13 \\
Syst. Descr. & Vehicle system repaired & Brakes \\
Job Location & Location of job completion & CODRF \\ \bottomrule
\end{tabular}
}
\end{table}

Together, these tables form a detailed, job-level dataset of maintenance on Detroit's entire vehicle fleet across 87 different departments, such as police, airport, fire, and solid waste. The records in each table are entirely complete (no fields are missing in any record). The data is, however, prone to noise, as often manually recorded by vehicle technicians at maintenance time (e.g., odometer readings fluctuated and sometimes even decreased between repairs) or ``lifetime to date'' statistics such as fuel consumption; hence there are potential concerns about the accuracy of some data due to human data-entry, job categorization errors, or data omitted from the electronic records. To minimize the impact of these uncertainties and utilize the most reliable data, following the recommendation of experts who are familiar with the data, we limit our analysis to maintenance records from 2010 or later, as Detroit's fleet data collection practices changed in 2010 (new electronic record-keeping system). This represents 1,087 active vehicles and over 25,000 maintenance records.

\section{Automated Multivariate Sequence Analysis with PRISM}
\label{sec:parafac}

We begin by addressing (RQ1): \textit{how can we uncover, validate, and interpret complex, multivariate patterns from fleet maintenance records?} Our aim is to identify meaningful multivariate maintenance patterns in the Detroit vehicle fleet, and to do so in a way that requires minimal human input and tuning so as to enable ongoing, automated analysis of maintenance event streams. We carefully design an algorithm that satisfies the following conditions: (i) the model is capable of extracting meaningful patterns from the fleet data with minimal tuning, (ii) the output is interpretable for a layperson, and (iii) the practitioners in the city can readily run the model when new data become available, needing minimal user intervention. To meet these requirements, we utilize PARAFAC as the foundation of this analysis, and then develop a novel algorithm, \textsc{PaRafac}-Informed Sequence Mining (PRISM), to identify ``characteristic subsequences'' unique to multivariate groupings identified by PARAFAC. PRISM assists in making the multidimensional patterns revealed by PARAFAC interpretable and actionable when applied to sequential data.

\subsection{Methodology}  

\subsubsection{Data Model}
\label{sec:datamodel}
\begin{figure}[!t]
    \centering
    \includegraphics[width = \columnwidth]{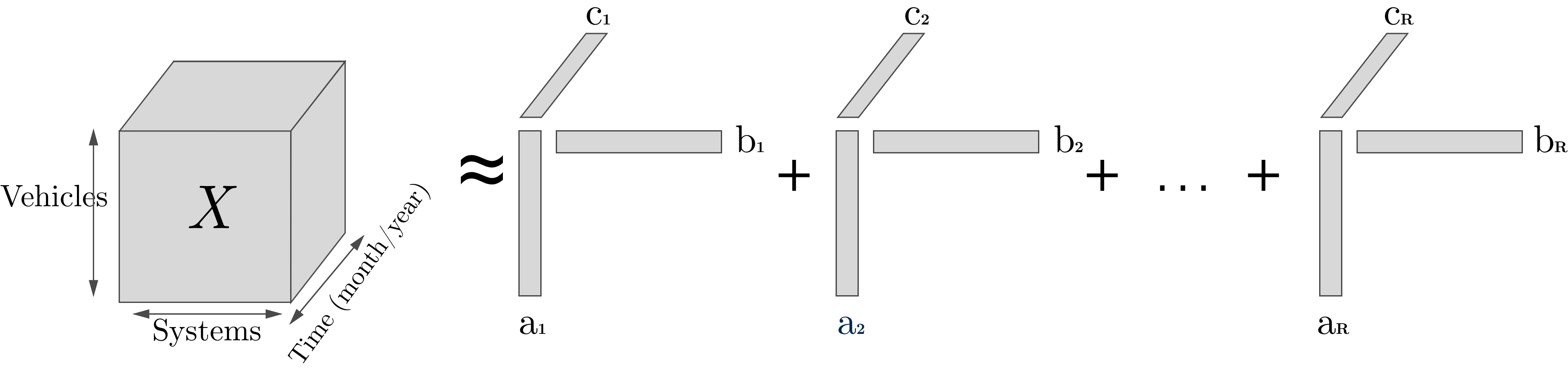}
    \caption{PARAFAC decomposes a $vehicle \times system \times time$ tensor into products of vehicle, system, and time factor vectors.} 
    \label{parafac-fig}
\end{figure}

Our goal is to encode the information of the entire fleet into a single dataset that will enable the discovery of meaningful fleet-level patterns. The multidimensional data described in \S~\ref{sec:data} can be naturally represented as tensors, or $n$-way arrays~\cite{Kolda2009-jh}. Specifically, we model  
the Detroit vehicle maintenance dataset as $vehicle \times system \times time$ data tensors. An illustration of a resulting 3-way tensor is shown in Figure~\ref{parafac-fig}, where the vertical axis (the \textit{first mode}) represents each different vehicle, sorted by year and unit number; the horizontal axis (the \textit{second mode}) represents each distinct vehicle system (see ``System Description'' in Table \ref{maintenance-table-supplementary}); and the depth (\textit{third mode}) represents time in months or years. The value at any given $[vehicle, system, time]$ entry in the tensor is the count of maintenance jobs for that particular vehicle, system, and time.

We note that in our data representation 
we do not attempt to separate different vehicle types and analyze them independently, 
as this type of user intervention drifts away from 
our goal of a fully automated data analysis pipeline. Most importantly, by grouping vehicles, there could be loss of information at the fleet level. A well-behaved algorithm should be able to find patterns at both the type- and fleet-level. In the following subsections, we demonstrate that both kinds of patterns are discovered through PARAFAC + PRISM.

\subsubsection{PARAFAC Decomposition}
\label{sec:parafac-decomp}
The PARallel FACtors (PARAFAC) decomposition is a higher-dimensional analog to the SVD, used for tensors in $> 2$ dimensions \cite{Kolda2009-jh}. 
PARAFAC decomposes a tensor into a sum of component rank-one tensors which best reconstruct the original tensor. For example, given a 3-way tensor $\mathcal{X} \in \mathbb{R}^{I \times J \times K}$, 
PARAFAC decomposes the tensor as $\mathcal{X} \approx \sum_{r=1}^R \mathbf{a}_r \circ \mathbf{b}_r \circ \mathbf{c}_r,$ where $\mathbf{a}_r \in \mathbb{R}^I$, $\mathbf{b}_r \in \mathbb{R}^J$, $\mathbf{c}_r \in \mathbb{R}^K$ for $r = 1, \ldots, R$  and ``$\circ$'' represents the vector outer product. The PARAFAC decomposition can be written compactly as the combination of three loading matrices \textbf{A}, \textbf{B}, \textbf{C}:
$\mathcal{X} \approx [\mathbf{A}^{I\times R},  \mathbf{B}^{J\times R},  \mathbf{C}^{K\times R}],$ in which the $r^{\rm th}$ columns correspond to the vectors $\mathbf{a}_r$, $\mathbf{b}_r$ and $\mathbf{c}_r$, respectively. These encode the most ``important'' relationships between different dimensions (or modes) of the tensor. For more information about PARAFAC, see \cite{Kolda2009-jh, lee2001algorithms}; details on our PARAFAC experiments are given in \ref{sec:appx-parafac}. 

The key aspect of the PARAFAC decomposition that makes it useful for understanding the Detroit vehicle-maintenance dataset is that it 
identifies $R$ groupings (factors) of different vehicles, systems, and times, as well as factor loading vectors $\mathbf{a}_r$, $\mathbf{b}_r$ and $\mathbf{c}_r$ which identify how strongly each \textit{vehicle}, \textit{system}, and \textit{time} contributes to this factor.

\subsubsection*{Limitations of PARAFAC}

There are several limitations to using PARAFAC alone to identify multivariate patterns:

\textbf{(a)} PARAFAC does not identify the individual \textit{observations} in each factor. PARAFAC only yields $R$ multivariate loading vectors $\mathbf{a}_r$, $\mathbf{b}_r$ and $\mathbf{c}_r$ indicating the degree to which each factor correlates with each index along each mode of the data. It is not clear how to utilize this information in downstream analysis beyond visualization of these vectors directly, as in Figures \ref{fig:month_year} and \ref{fig:vehicle_year}. As a result of this limitation, we cannot answer the question: to which $[vehicle, system, time]$ observations does factor $r$ apply (or not apply)? This prevents, for example, searching for vehicles or maintenance records falling under a specific factor. As a result of this limitation, we cannot provide technicians with a list of vehicles in a specific PARAFAC factor for further inspection or repair, nor can we compute the total cost of maintenance within a given PARAFAC factor to share with fleet managers or policymakers. 

While sparsity-inducing PARAFAC decomposition algorithms exist, in this application, we do not actually have prior knowledge that the underlying structural relationships are indeed sparse. Imposing sparsity constraints may lead to incorrect conclusions. Vehicle maintenance data reflects complex relationships between vehicles, systems, and time, which may not match the assumptions of a sparsity-inducing PARAFAC. Instead, we desire a solution which imposes minimal assumptions on the data while still allowing for inference about the in- and out-groups in each resulting factor component for downstream analysis.
    
\textbf{(b)} PARAFAC does not directly leverage the sequential nature of the data. PARAFAC only uses the \textit{frequency} of $[vehicle, system, time]$ triplets in the data tensor. Due to this limitation, we cannot identify the specific \textit{sequences} from the underlying data that give rise to the high loadings in each factor $r$, and cannot answer the question ``what observed maintenance subsequences in the original data give rise to factor $r$?'' As an example, the PARAFAC loading vectors would not differentiate between the sequences ``Accident, Brakes, Brakes`` and ``Brakes, Brakes, Accident'', but these sequences lead to different hypotheses about underlying fleet maintenance issues in a factor grouping (the first implies accidents frequently result in brake damage; the second implies brake issues frequently precede accidents). 

Extracting these sequences requires manual interpretation of the results, which can be both labor-intensive and ad hoc: users must attempt to discern which vehicles, systems, and times each factor applies to (using three-way plots), and then undertake a separate analysis of the repair sequences for those vehicle-system-time combinations. 

There is no existing methodology to address this limitation of PARAFAC for sequential data, despite the fact that many previous applications of PARAFAC also evaluate data which is sequential in nature (e.g. text \cite{Bader2008-rm} and discourse \cite{Acar2006-yl} data).

\subsubsection{Differential Sequence Mining (DSM)}\label{sec:dsm}

Limitation (b) of PARAFAC could be addressed via differential sequence mining (DSM), which identifies differences in sequences between two groups. Existing methods for DSM rely on computing frequent sequences in a group of interest (which we refer to as the \textbf{``in-group''}), and comparing their frequency to another group (the \textbf{``out-group''}) using statistical tests. A common method for DSM computes the \textit{i-ratio}, $\frac{|InGroup|}{|OutGroup|}$, and uses a $t$-test to determine whether the observed i-ratio is statistically significant \cite{Kinnebrew2013-ny}.\footnote{In the original work, ``in-group'' and ``out-group'' are referred to as left and right groups, respectively, but the meaning here is the same.}

However, several limitations of existing DSM methods make it ineffective for the current application. First, DSM is only useful if the first limitation of PARAFAC is solved: the i-ratio requires a binary identification of whether each observation is ``in'' or ``out'' of a given PARAFAC factor. As mentioned above, the only methods to do so would require imposing sparsity constraints on the resulting decomposition, which we seek to avoid. Second, the frequent pattern search algorithm used in DSM is based on overall frequency, without regard to the ``uniqueness'' of those sequences to the in-group, and so yields little additional information. Third, its use of \textit{frequency} yields results which are biased toward shorter subsequences.  Finally, the extensive use of frequentist statistical significance testing in DSM \cite{Kinnebrew2013-ny}, where a $t$-test is applied to every subsequence evaluated, can lead to spurious results and ``statistically significant'' results which merely reflect large sample sizes, not large effect sizes \cite{Wasserstein2016-gm}. This is the case even when most commonly-used corrections for multiple hypothesis testing (e.g. Bonferonni, Benjamini \& Hochberg) are applied, as these are only appropriate for small numbers of tests \cite{Efron2016-to}, while thousands of subsequences are commonly evaluated in tasks such as our case study below. In the context of large-scale data analysis where many subsequences (e.g. all $n$-grams of length $\leq 5$) may be evaluated to compare many different subgroups, the Type I Error rate of such tests breaks down \cite{Efron2016-to}.

\begin{algorithm}[t!]
	\caption{PRISM: executed on the factor loading matrices of each of the $R$ PARAFAC factors.}
    \label{alg:prism}
	\begin{algorithmic}[1]
	
		\State \textbf{Input}: $\mathbf{a}_r, \mathbf{b}_r, \mathbf{c}_r$: loading vectors for factor $r$; seqs: list of vehicle maintenance sequences; and priors, $\gamma$, BDPT prior and ROPE.
		\State \textbf{Output}: $\Delta \theta_{seq}$: posterior difference in proportions; and $\mathbb{P}(\Delta \theta_{seq} \notin ROPE)$: probability of practical difference in proportions for all frequent sequences in in-group.
		\vspace{0.2cm}
		
		\State \textcolor{gray}{/* In practice the algorithm is not sensitive to the parameters of either BGMM (e.g. $\gamma$) nor to the choice of prior in BDPT as long as a weak, uninformative prior is used and $\gamma$ is not near the extremes of $[0,1]$.  */}
		\State initialization ($k=2, \gamma = \frac{1}{2}, ROPE=0.01$)
		
		\vspace{0.15cm}
		\State \textcolor{gray}{/* \textbf{S1}: Determine in-group observations per mode $\{a, b, c\}$ using a Bayesian Gaussian Mixture Model (BGMM). */}
		\ForAll { $LoadingMatrix$ in $\{a,b,c\}$ } 
             \State $\textrm{InGroup}_i \leftarrow$ BGMM($\textrm{LoadingMatrix}_r$, $\gamma=\frac{1}{2}$)
        \EndFor
        \vspace{0.1cm}
        \State \textcolor{gray}{/* \textbf{S2}: Find high-frequency sequences for the in-group vehicles. */}
		\State InGroupSeqs $\leftarrow$ Filter(seqs,  ($\textrm{InGroup}_a$)
        \State OutGroupSeqs $\leftarrow$ Filter(seqs, $\neg (\textrm{InGroup}_a$))
        \State m = $| \textrm{InGroupSeqs} |$
        \State n = $| \textrm{OutGroupSeqs} |$
        \State InGroupFreqSeqs $\leftarrow$ FindFreqSeqs(InGroupVehicleSeqs, $\textrm{InGroup}_a$)
        \vspace{0.1cm}
		\State \textcolor{gray}{/* \textbf{S3}: Conduct Bayesian Difference in Proportions Test (BDPT). */}
		\ForAll { seq in InGroupFreqSeqs} 
              \State $\textrm{InGroupSupp}
              \leftarrow \sum{\mathbbm{1}}_{\textrm{InGroupFreqSeqs = seq}}$
              \State $\textrm{OutGroupSupp}
              \leftarrow \sum{\mathbbm{1}}_{\textrm{OutGroupFreqSeqs = seq}}$
              \State [$\Delta \theta_{seq}, \mathbb{P}(\theta \notin \textrm{ROPE})_{seq}$] = BDPT(InGroupSupp,
              
              \hfill  OutGroupSupp, m, n) 
        \EndFor
	\end{algorithmic}
\end{algorithm}

\subsubsection{\textsc{PaRafac} Informed Sequence Mining (PRISM)}\label{sec:prism}

Motivated by our observations in \S~\ref{sec:parafac-decomp} and \ref{sec:dsm}, we present an algorithm, \textsc{PaRafac}-Informed Sequence Mining (PRISM), which jointly resolves the existing limitations of prior DSM algorithms and includes \textit{the first unified, automated approach} to link DSM to the results of a PARAFAC analysis. 
We give its pseudocode in Algorithm~\ref{alg:prism}. 
At a high level, it consists of the following steps for each PARAFAC component $r = 1 \ldots R$:
\begin{enumerate}
    \item[\bf S1] A Bayesian Gaussian Mixture Model (BGMM) is used to identify the ``in-group'' vehicles, systems, and time points for a factor $r$ (those to which this factor applies). We use a standard finite mixture model with $k = 2$ components, a Dirichlet distribution, and a standard weight concentration prior of $\gamma = \frac{1}{k} = \frac{1}{2}$, fit separately to each factor loading vector. The in-group for each dimension is the mixture component with a larger posterior mean. In practice, this procedure  separates observations with near-zero and non-zero entries in $\mathbf{a}_r$, $\mathbf{b}_r$ and $\mathbf{c}_r$ quite effectively, without much sensitivity to $\gamma$. We give more details in App.~\ref{sec:appx-bgmm}.
    \item[\bf S2] Compute frequent sequences for the in-group vehicle-system-time set using a standard frequent sequence mining algorithm \cite{Wang2007-qp}, and only keep sequences which contain at least one in-group system. Normalize frequencies by the total size of each group (i.e., total number of $n$-grams in in-group and out-group, respectively) to produce a proportion. 
    \item[\bf S3] Conduct a Bayesian difference-in-proportions test (BDPT) using a non-informative prior (e.g., $\textrm{Beta}(1,1)$, the weakest form of the conjugate prior for a binomial proportion) to determine the posterior probability of whether the proportion of the observed subsequences in each group is the same. The resulting subsequences for which the posterior probability of a large difference in proportions between in-group and out-group vehicles is below some predetermined threshold (e.g., $0.05$) are the ``characteristic subsequences'' of that factor. Replication details are given in App.~\ref{sec:appx-bdpt}.
\end{enumerate}

PRISM thus jointly resolves the limitations of PARAFAC described above. \textbf{S1} determines, for every $[vehicle, system, time]$ maintenance record, whether it is ``in'' factor $r$ or not. Then, \textbf{S2} mines the ``in-group'' for factor $r$ to determine which maintenance sequences, for those $[vehicle, system, time]$ records in the factor, are most unique to factor $r$. \textbf{S3} ensures the identified sequences are both statistically significant and practically important by ensuring that the posterior probability that the difference in proportions is larger than ROPE is high, according to BDPT.

\begin{figure*}[t!]
    \centering
    \includegraphics[width=\textwidth]{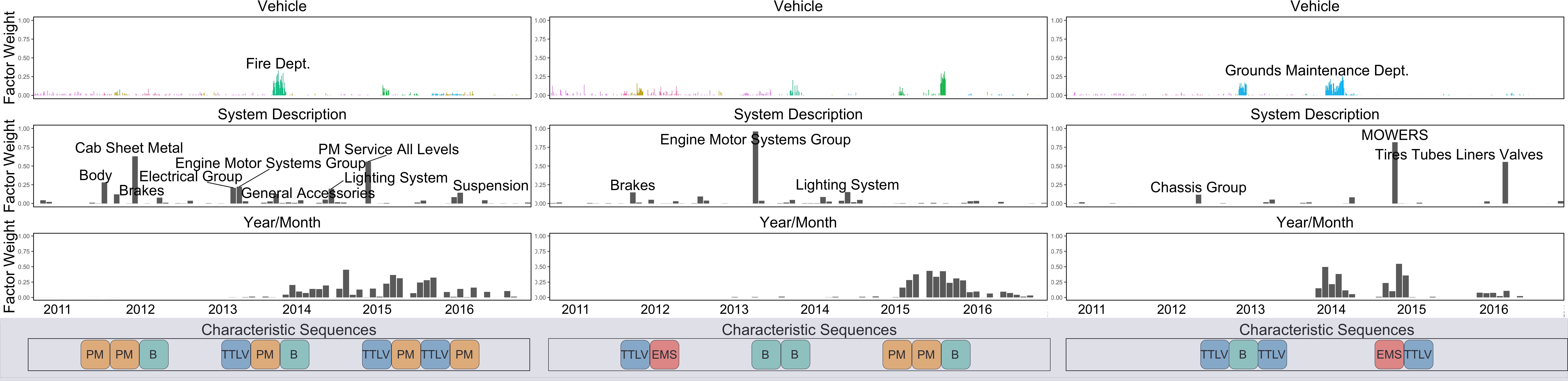}
    \caption{Top white Panel: PARAFAC 3-way plot of \textit{absolute-time} analysis. Patterns involving the highlighted vehicles (top row) going under specific types of repairs (middle row) over select times (bottom row) are shown. Left column: Ambulance 2014 Terrastar Horton vehicles involved in Body (B), Cab/Sheet Metal, Engine and Motor (EMS), and Preventive Maintenance (PM) services after 2014. Center column: Repair to specific systems of the Smeal SST Pumper (fire truck), from late 2015 through 2016. Right column: System and time patterns for riding mowers, with repairs to mower blades and tires/tubes/liners/valves (LLTV) during seasons of high usage. Bottom gray Panel: A subset of the characteristic maintenance subsequences discovered via PRISM applied to the corresponding factor vectors.
} 
    \label{fig:month_year}
\end{figure*}

\begin{figure*}[t!]
    \centering
    \includegraphics[width = \textwidth] {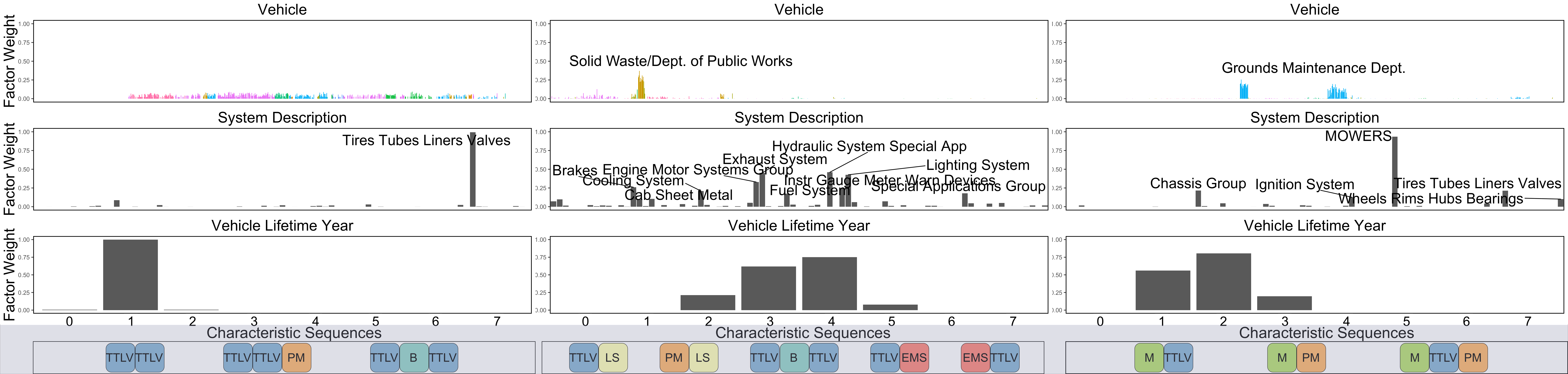}
    \caption{Top white panel: PARAFAC 3-way plot of \textit{vehicle lifetime} analysis. Left column: Simple pattern common to almost \textit{all} vehicles: tires/tubes/valves/liners (TTLV) replacement during the second year of lifetime. Center column: The 2012 Freightliner M2112V, a garbage truck, has increased maintenance in years 2-4 after purchase, focusing on hydraulics, lighting (LS), gauges and warning devices, and cooling systems. Right column: Patterns primarily for the 2013 Hustler Z 60 2013 (a riding mower), which have mowing blades (M) serviced frequently in the second and third years of their lifetime. Bottom gray panel: A subset of the characteristic maintenance subsequences discovered via PRISM applied to the corresponding factor vectors.}
    \label{fig:vehicle_year}
\end{figure*}

PRISM provides a unified method for leveraging the valuable data provided by the PARAFAC factor loading matrices \textbf{A}, \textbf{B}, \textbf{C} via sequence mining in order to identify ``characteristic subsequences'' specific to the multidimensional loadings of each factor $r$.  
This information is not given by PARAFAC alone. Furthermore, using a Bayesian framework for both the clustering and, in particular, the statistical analysis of subsequences in DSM alleviates concerns about multiple hypothesis testing, as each iteration is simply estimating the posterior probability of a difference in relative frequency between the in- and out-groups, \textit{not} the probability that we would observe the data due to random chance under $H_0$, which would require controlling for Type I Error \cite{Gelman2012-ys}. Additionally, instead of simply evaluating a point hypothesis (typically $H_0: \theta_{in} = \theta_{out}$), the Bayesian test allows us to estimate the probability that the difference in frequencies is outside of a ``region of practical equivalence'', or ROPE \cite{Kruschke2011-cz}, which excludes what might otherwise be ``statistically significant'', but practically useless, results in the case of small but genuine differences in frequency of occurrence. We discuss uses of such sequences in Section \ref{sec:parafac-results-impact}.

\subsection{Findings and Impact}\label{sec:parafac-results-impact}

\subsubsection{PARAFAC} \label{sec:results-parafac} \textbf{Setup.} There is no explicit methodology of which we are aware for selecting $R$. In our analysis we set $R = 25$, but the results that we report are largely robust to different values of $R$. Our choice is consistent with the literature (see \S~\ref{sec:related}) and also leads to a manageable number of 3-way plots ($2 \times 25$ factors per our analysis)  that can be easily inspected by a civic data scientist. Details on the objective function, algorithm, and convergence of the PARAFAC model used here are given in \S \ref{sec:appx-parafac}. 

First, we seek to identify multivariate vehicle-system-time relationships in the Detroit dataset in a way that is automated and interpretable, even for non-technical domain experts and city stakeholders. To this end, we generate ``3-way'' plots of the three factor matrices from the PARAFAC decomposition \cite{Koutra2012-hx} using the tensor toolkit provided by \cite{Bader2007-si, Bader_undated-ue}, as shown in Figures~\ref{fig:month_year}-\ref{fig:vehicle_year} (top, white panels). Each plot visualizes the vectors $\mathbf{a}_r$, $\mathbf{b}_r$ and $\mathbf{c}_r$, which show the different modes (vehicle, system, time) participating in the $r^{\rm th}$ factor.  We explore two different representations of time in the data tensors: one which uses \textit{absolute time} (month and year) in Figure \ref{fig:month_year} and another using \textit{vehicle lifetime} (by year, starting with the vehicle's purchase year) in Figure \ref{fig:vehicle_year}. The absolute time analysis allows us to model seasonality and other real-time trends in fleet maintenance, and could be more useful in forecasting future maintenance. On the other hand, the vehicle lifetime analysis allows us to measure trends and changes in vehicles' maintenance over the course of their lifetime in the Detroit fleet, and could be useful for vehicle reliability analyses. 

\vspace{0.1cm}
\noindent \textbf{Findings.} Examples of the results from the absolute time analysis are shown in Figure \ref{fig:month_year}. These results demonstrate clear patterns across vehicles, systems under repair, and time, underscoring the importance of this multivariate approach. For example, fire trucks and ambulances (the Terrastar Horton in left column of Figure \ref{fig:month_year} and Smeal SST Pumper in the center column of Figure \ref{fig:month_year}, respectively) both show strong evidence of patterns in their maintenance, but with very different groups of systems and across different time bands. The riding mower shown in the right column of Figure  \ref{fig:month_year}, however, displays an entirely different maintenance pattern, with a focus on only two systems (mowing blades and tires/tubes/liners) and strong seasonality, which reflects the seasonal use of mowers in a northern city such as Detroit.

Examples of the results from the PARAFAC vehicle lifetime analysis are shown in Figure \ref{fig:vehicle_year}. This analysis demonstrates a different set of patterns: those across the lifetime of vehicles, beginning when they are purchased. Note that the right column of Figures \ref{fig:month_year} and \ref{fig:vehicle_year} identify a nearly identical set of vehicles but highlight different patterns, illustrating the different insights gained from absolute time vs. lifetime analyses. Additionally, the center and right columns of Figure \ref{fig:vehicle_year} are an examples of vehicle-level maintenance patterns, while the left column of Figure \ref{fig:vehicle_year} is an example of fleet-level maintenance patterns which is common across the entire fleet. This example illustrates that PARAFAC is indeed capable of automatically discovering patterns at both vehicle and fleet level, as desired (\S~\ref{sec:datamodel}). 

Figures \ref{fig:month_year} and \ref{fig:vehicle_year} show how patterns specific to certain \textit{departments} are automatically uncovered by PARAFAC, even though departmental data was not provided in the input data to PARAFAC. We later also learned that the factors in Figures \ref{fig:month_year} relating to ambulance and fire trucks were actually indicative of specialist technicians working on those vehicles; again, PARAFAC revealed these unique multidimensional patterns without preexisting knowledge.

\subsubsection{PRISM} 

\textbf{Setup.} The PRISM algorithm allows us to leverage the PARAFAC loadings to extract further insight about each group, by mining sequences which represent specifically the vehicle/system/time observations represented in each factor's loading vectors $\mathbf{a}_r, \mathbf{b}_r, \mathbf{c}_r$. This analysis uses $ROPE = 0.01$, i.e., PRISM searches for subsequences which have high posterior probability of differing in normalized frequency by at least $0.01$ between the in- and out-groups of any given factor according to BDPT (in most cases, the observed difference is much larger). In Figure \ref{fig:month_year} and Figure \ref{fig:vehicle_year}, we add a subset of the characteristic maintenance subsequences discovered via PRISM applied to the corresponding factor vectors. These are shown in the bottom gray panel below each three-way plot. The specific characteristic sequences presented here were selected from a larger set of overall PRISM results for each factor. 

\vspace{0.1cm}
\noindent  \textbf{Findings.} The sequences identify concrete vehicle repair sequences which are uniquely common to the vehicle/system/time grouping in each factor. For example, we might use the characteristic sequences to recommend brake service (B) whenever preventive maintenance (PM) is performed for the vehicles in the factors in the left and center columns of Figure \ref{fig:month_year} (mostly ambulance and fire truck), or to recommend lighting system repairs when PM is performed for vehicles in Figure~\ref{fig:vehicle_year}b (garbage truck). Furthermore, PRISM provides validation of the PARAFAC loadings, confirming that there are significant differences in the occurrence of maintenance patterns across the vehicle/system/time groups identified via PARAFAC.

\subsubsection{Impact}

The PARAFAC + PRISM analysis demonstrates the variety of insights that can be gained from using tensor decomposition to understand multidimensional data. The analysis above uncovers multidimensional patterns across the entire Detroit vehicle fleet, as well as unique trends specific to certain vehicles, systems, and times. Additionally, the use of two different measures of time---month/year, and vehicle lifetime---allows us to demonstrate two different modes of time-bound pattern in the data. These results suggest several potential actions for Detroit, including potential seasonal allocation of resources and technicians (e.g., for mower system repair during the summer time, as shown in the right column of \ref{fig:month_year}), and point to future efforts in detailed analyses of such data for other purposes, such as anomaly detection and automated fleet maintenance recommendation or scheduling systems.

The PRISM algorithm provides, to our knowledge, the first principled method to automatically extract interpretable information from the results of PARAFAC and utilize it for sequence analysis. It has the potential to apply more broadly to a variety of sequence mining tasks where the unsupervised identification of groups and their defining sequential patterns is desired. PRISM can specifically inform future work on predicting vehicle maintenance, availability, and labor, parts, and other costs due to maintenance. It could also potentially lead to changes in the city's fleet maintenance operations by providing interpretable visualizations to policymakers and vehicle mechanics, as well as providing suggested maintenance ``bundles'' for individual vehicles or groups of vehicles while they are in for repair, which could lead to economies of scale and improved cost efficiency as the city works to emerge from its bankruptcy.
Moreover, our methodology generalizes to other domains where multidimensional, sequential data abound, including tasks to which PARAFAC has been previously applied (see \S \ref{sec:related}).

\section{Forecasting Maintenance Patterns}\label{sec:lstm}

Our results in \S \ref{sec:parafac} demonstrate the existence of vehicle-system-time maintenance patterns which could be exploited by appropriate sequence models in order to address additional needs. Our task in this section is to leverage these patterns build a set of \textit{predictive} models for a \textit{specific type} of vehicles, unlike \S~\ref{sec:parafac} where our task was to uncover sequential maintenance patterns from the \textit{entire} dataset. Specifically, we address (RQ2), \textit{Can we predict vehicle maintenance?}, and (RQ3), \textit{Can we predict vehicle- and fleet-level maintenance costs?}. (RQ2) deals with the low-level details of maintenance prediction, and (RQ3) is a high-level prediction task that is critical for budgeting in large, financially-strained municipalities such as Detroit.

To address these questions, we construct two models, one for each task, that predict the next item (maintenance job or maintenance costs) in a time series for vehicles in the fleet, given a set of previous items. We illustrate that simple, standard models achieve good performance, implying that these tasks are highly amenable to data mining.  

\vspace{0.1cm}
\noindent \textbf{Data.} Per our stakeholders' request, in this section we focus on Detroit's police vehicles, consisting of Dodge Chargers, Chevrolet Impalas, and Ford Crown Victorias. Police vehicles, particularly in a large and budget-strained city such as Detroit, are  critical to the city's capacity to deliver services, and represent a substantial portion of vehicle usage, maintenance, and procurement costs. Using these vehicles as a case study allows us to focus on identifying, modeling, and interpreting patterns specific to police vehicles, while also demonstrating the broader potential of our methods' ability to answer the specified questions for other vehicles in future analyses, or leveraging our open-source code for analysis of other domains. 
\subsection{Methodology}

\subsubsection{Maintenance Sequence Forecasting}

We implement a sequential model to predict vehicle maintenance using the sequential structure of maintenance patterns (\S~\ref{sec:parafac}), which can be useful for resource allocation, technician hiring, or the preparation of a data-driven budget proposal. 
Specifically, we utilize the Long Short-Term Memory (LSTM) neural network \cite{Zaremba2014-vx}, a well-established model that reads over a sequence, one item at a time, and computes probabilities of the possible values for the next item in the sequence. In theory, an LSTM is capable of learning long-distance dependencies across a sequence \cite{Hochreiter1997-jm}. 

\vspace{0.1cm}
\noindent \textbf{Data Setup.} From the raw data, we assemble a dataset consisting of the complete sequence of system repairs for each vehicle. Each vehicle's sequence is considered a separate observation. To assemble training, validation, and testing datasets for the model, we use all data from the three vehicles predominantly used as police cars in the Detroit fleet. Ideally, a model would be fit on only a single vehicle type; however, due to the relatively small number of vehicles available for training (329 total police vehicles), it was necessary to combine multiple make/models. We train on a random subset of 50\% of vehicles, using 25\% for model validation and 25\% for testing.

\vspace{0.1cm}
\noindent \textbf{Evaluation.}  An effective model assigns high probability to unseen data and low probability to a repair job that does not happen. Hence, we choose to assess the performance of our model using average per-item perplexity, a common evaluation metric for sequence models which evaluates the probability assigned to entire test sequences: $e^{-\frac{1}{N} \sum_{i=1}^N\ln(p_{{\rm target}_i})} = e^{\rm loss}$, where $N$ is the total number of observations and $p_{{\rm target}_i}$ is the probability assigned to item $i$. Assigning a high probability to true, unseen data is equivalent to achieving low perplexity.

\begin{figure}[t!]
    \centering
    \includegraphics[width = .92\columnwidth]{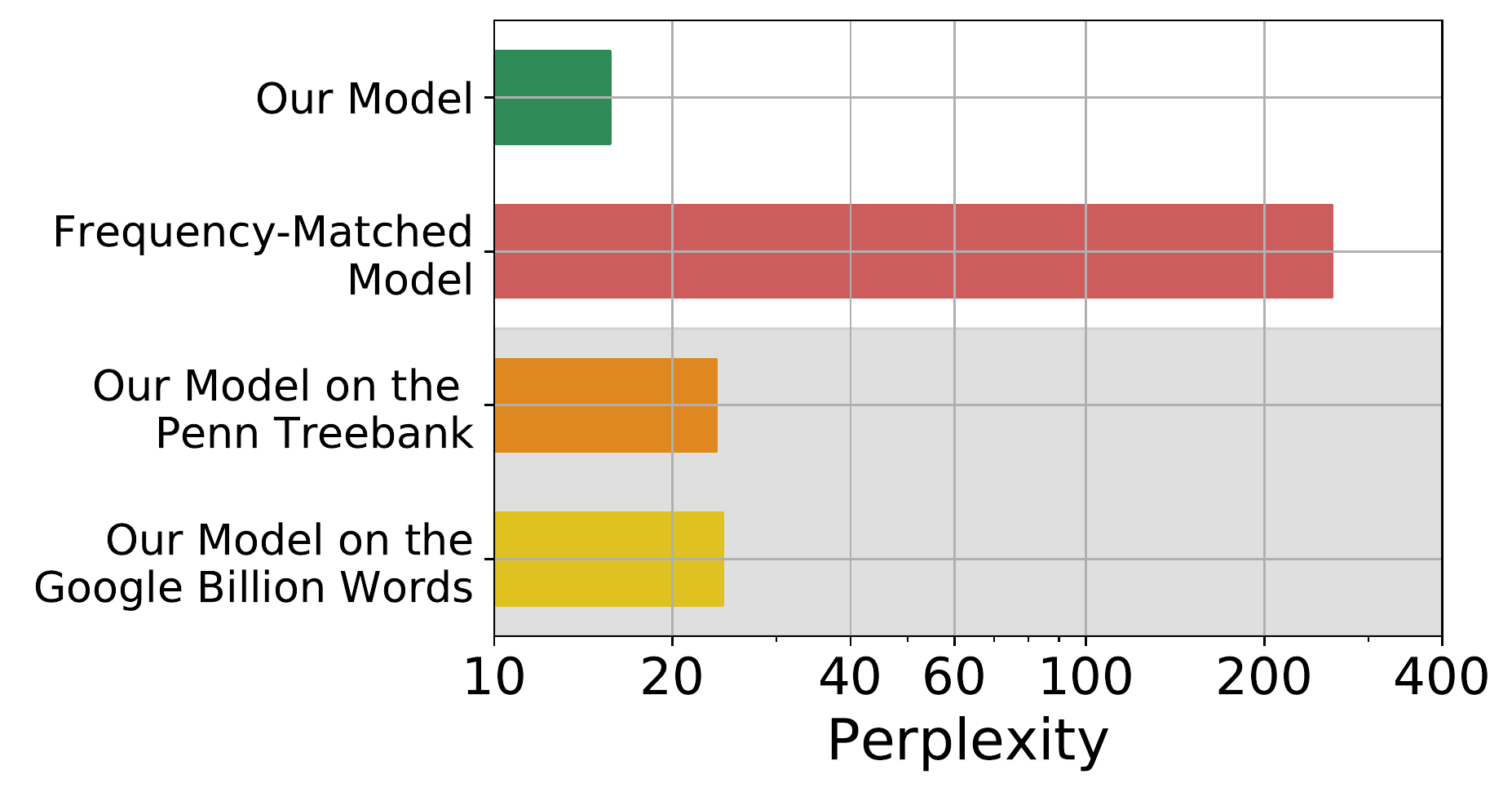}
    \caption{
    Performance of our model in predicting the probability of the next maintenance job in a sequence (green) vs. a frequency-matched model (red), plus the performance of our model on external datasets (orange and yellow).}
    \label{perplexity-fig}
\end{figure}

\textit{Baselines.} We compare the LSTM model to a baseline that we call \textit{frequency-matched model}. In this model, we first compute the frequency of item $i$ over all sequences in the training data. Then we use this frequency to assign a probability to each target observation in the test sample, $p_{{\rm target}_i}$, and compute the perplexity score. Because there are no other maintenance prediction models in prior published work, we also provide the perplexity score of our model on two external datasets. These results, along with the results of our model, are shown in Figure \ref{perplexity-fig}.

\vspace{0.1cm}
\noindent \textbf{Model.} We implement the well-known LSTM architecture originally used in  \cite{Zaremba2014-vx} because of its ability to model complex sequences while avoiding overfitting. The model is a 2-layer LSTM which reads over maintenance sequences in temporal order, maintaining a window size of at most 20 observations. Detailed training hyperparameters are given in \S \ref{sec:appx-lstm}.

\subsubsection{Maintenance Cost Forecasting}

We forecast maintenance costs for active police vehicles using an autoregressive integrated moving average (ARIMA) model. Recent work has demonstrated that ARIMA performs well even in comparison with other highly complex machine learning methods for time series data \cite{Makridakis2018-ty}. 
Moreover, it well-known theoretical properties and interpretability make it ideal for our analysis. 

\vspace{0.1cm}
\noindent \textbf{Data Setup.} All of our forecasts are in terms of average monthly cost per vehicle. The cost data includes frequent fluctuations caused by decommissioning and acquiring vehicles (see Figure \ref{arima-fig}, which makes the prediction task challenging. We use a monthly timescale as a balance between aggregating enough data per time period to be sufficiently stable and detecting variation on smaller timescales (e.g., seasonality). 

\vspace{0.1cm}
\noindent  \textbf{Evaluation.} The forecast model is evaluated using predictions of costs one and six months into the future. We evaluate the model using its root mean squared error (RMSE), but we also monitor AIC and BIC during model fitting in order to select hyperparameters.

\vspace{0.1cm}
\noindent  \textbf{Model.} Our models predict the average cost of an entire \textit{department} (police), or the average cost of a specific \textit{make/model} (Dodge Charger, Crown Victoria). Each ARIMA model is trained on data from the first 24 months, and generates predictions of the average cost per vehicle. Predictions are made one month and six months into the future. The model is then updated with the true average cost per vehicle from the 25$^{\rm th}$ month, and generates the next pair of forecasts. This is a standard training regime for autoregressive time series models. For the details of model training and final ARIMA hyperparameter settings, see \ref{sec:appx-arima}.

\subsection{Findings and Impact} 

\subsubsection{Maintenance Sequence Forecasting}

Figure~\ref{perplexity-fig} compares the performance of our LSTM model with the frequency-matched model in predicting the next item in a maintenance sequence on the Detroit dataset. We also present the performance of the same model on external datasets. Our model achieves an average test perplexity score of 15.7, demonstrating that even this relatively simple, computationally lightweight model with a small dataset is able to achieve strong predictive performance, far better than the frequency-matched model's perplexity of $260 \pm 40$. 

For comparison, we note that the architecture used here has also achieved perplexity score of 23.7 on the Penn Treebank dataset and 24.3 on the Google Billion Words dataset \cite{Kuchaiev2017-ip}. While our model's low perplexity score should not be directly compared to model performance on other corpora, because of the relatively low number of candidate items in the sequence -- 81 unique systems in the entire vehicles dataset compared to many thousands in text corpora -- the reference indicates that our model assigns probability scores with performance on par with state-of-the-art language models.

\subsubsection{Maintenance Costs Forecasting}

\begin{figure}[t]
    \centering
        \includegraphics[width = .49\columnwidth]{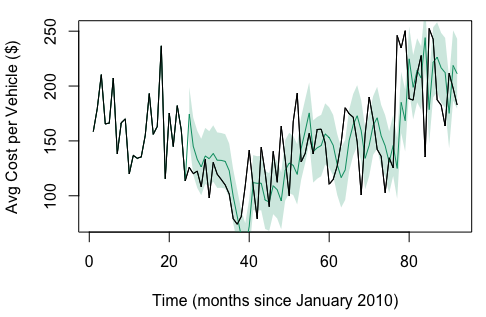}
    \includegraphics[width = .49\columnwidth]{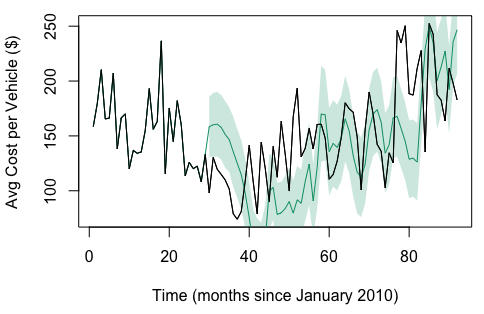} \\
    \includegraphics[width =.49\columnwidth]{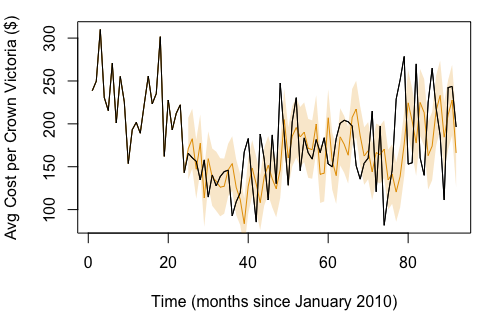}
    \includegraphics[width = .49\columnwidth]{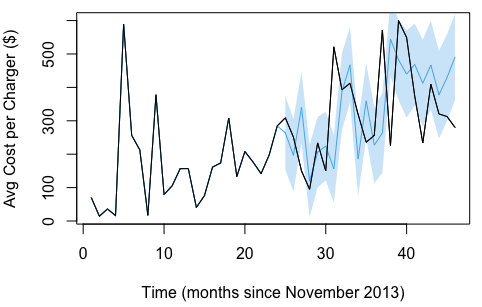}
    \vspace{-0.3cm}
    \caption{Top: One-month (left; RMSE = \$38.6) and six-month (right; RMSE = \$49.3) cost forecasts for police department. Bottom: One-month cost forecast for police vehicles by model, Ford Crown Victorias (left; RMSE = \$49) and Dodge Chargers (right; RMSE = \$158). 68\% confidence intervals shown. Ground-truth costs shown in black.}
    \label{arima-fig}
\end{figure}

Figure~\ref{arima-fig} shows the results of the cost forecasting models, along with the ground truth costs. The models show good agreement with the actual observations. For the department-level model (top of Figure~\ref{arima-fig}), the RMSE in predicting average per-vehicle cost ranges from \$38 to \$49, increasing only gradually as the prediction distance increases from 1 to 6 months, suggesting that the model is capable of making both short-term and medium-term predictions.
For the vehicle-specific model (bottom of Figure~\ref{arima-fig}), we show that the model is able to forecast costs for Ford Crown Victorias and Dodge Chargers. The Charger prediction is particularly challenging given the small sample and the rapid fluctuation due to new Charger acquisitions during the period of analysis.

\subsubsection{Impact}

Our analysis indicates that it is possible to accurately predict both future maintenance jobs and the average future expenses, both of which are critical for planning purposes.
Specifically, we show that future vehicle maintenance sequence can be predicted with high accuracy even in a modestly-sized fleet (164 training observations). The predictions of the LSTM can be used, for example, to support automated maintenance scheduling, availability or cost forecasting based on maintenance predictions, dynamic allocation of technicians and budget, anomaly detection, and many other applications which can ensure effective fleet-wide maintenance.

Moreover, our vehicle- and department-level cost models demonstrate that relatively accurate per-vehicle cost predictions (e.g. within 20-25\% at the department level for predictions one and six months into the future) can be obtained using a simple model and only 24 months of prior data---a historical window which any municipality should have available. These models can support budgeting and cost projection for data-driven planning, as well as comparative analysis of the current and projected future per-vehicle costs of different vehicle models. Cost projections are important for informing future purchasing, maintenance, usage, and vehicle disposal decisions. They can also contribute to optimal fleet composition prediction, which can allow Detroit to optimize the vehicles deployed for achieving service delivery and cost goals. Such tasks can be particularly impactful as the city recovers from bankruptcy.

Our analysis shows that even simple models (such as ARIMA) have significant predictive power for vehicle fleet analysis tasks.
Future directions include utilizing the output of the LSTM model in order to potentially further improve the accuracy of ARIMA.

\section{Related Work}\label{sec:related}

Our analysis is based on tensor decomposition and related to studies on municipal vehicle fleets and municipal forecasting.

\vspace{0.1cm}
\noindent \textit{Tensor Analysis and Applications.} Tensor representations and various decompositions have found wide applications in a variety of domains, including psychometrics \cite{Douglas_Carroll1970-jg}, epidemiology \cite{Sakurai2015-iz},
modeling online discourse over time \cite{Bader2008-rm, Acar2006-yl}, web search \cite{Sun2005-dm}, and anomaly detection \cite{Koutra2012-hx}. For a more detailed overview of tensor decompositions see \cite{Kolda2009-jh}.

\vspace{0.1cm}
\noindent \textit{Municipal Vehicle Fleets Research.} While predictive analytics, data science, and their application to urban planning (also known as \textit{urban informatics}) have dramatically expanded in recent years, these techniques have seen only limited applications to one of the largest and most substantial assets managed by many governments---their vehicles---and published research on the topic is surprisingly limited. Some state and local governments conduct, but rarely publish, fleet lifecycle reports and maintenance analyses \cite{Gransberg_undated-nb} and fleet management \cite{Osborne2012-ws, Lauria2014-ph} mostly focused on cost reduction.

Research on predictive maintenance has utilized on-board vehicle data for maintenance prediction \cite{Prytz2014-na} and for evaluating winter maintenance \cite{Lee2008-yz}. There have been some applications of deep learning to vehicle data for e.g. identifying faulty components and vehicle damage from photos \cite{singh2019deep}, but no prior work on mining or modeling fleet maintenance records. Other vehicle-related issues in urban areas have received significant research attention, including accident prediction \cite{Levine1995-cr} and traffic flow prediction and optimization \cite{Vlahogianni2005-yn, Weizhong_Zheng2006-sf, Lv2015-li}. The authors are not aware of any prior research applying tensor decomposition or the other techniques used in the current work to municipal vehicle data.

\vspace{0.1cm}
\noindent \textit{Municipal Forecasting.}  
Prior work has explored forecasting tasks in other areas of municipal government, including predictions of water usage \cite{Campisi-Pinto2012-ll} and solid waste generation \cite{Johnson2017-rv}. Prior work has also examined the use of decision support systems utilizing ARIMA and other time series models \cite{Rego2015-al}, but budgetary forecasting is still widely considered an open problem in municipal government, largely due to the complexity of the interests and constraints involved \cite{Forrester1991-un}.

\section{Conclusions and Discussion}
\label{sec:conclusions}
In this analysis, we describe the results of a data collaboration with Detroit's Operations and Infrastructure Group. This work applies methods to uncover maintenance-related patterns relevant to three key research questions. Our key contribution is to extract multidimensional maintenance patterns across the entire fleet using PARAFAC and the PRISM algorithm, which identifies characteristic subsequences for each PARAFAC factor (RQ1). We emphasize that the output of the PARAFAC algorithm is hardly interpretable. To alleviate this shortcoming, we propose the PRISM algorithm that can extract interpretable results from PARAFAC factors. We then move on to predictive tasks, one low-level and one high-level.
We build an accurate maintenance forecasting model which predicts the next maintenance job using fewer than 200 vehicles for training (RQ2). We conduct maintenance cost forecasting at department- as well as individual-level vehicle (RQ3). We show that even simple, standard, highly-interpretable predictive models achieve good performance and provide actionable insights to our partners in the City.

To the best of our knowledge, this work provides the \textit{first data-driven baseline for future studies} on applying data mining to municipal vehicle data. We set a precedent in this domain and publicly release our code to enable other cities and organizations to replicate or extend this analysis on their own fleet data.

\vspace{0.1cm}
\noindent \textbf{Limitations.} As all empirical studies, our analysis has some limitations. We highlight areas where our analysis was limited by data issues, and where future practitioners and analysts ought to direct data collection efforts. Future data collection efforts should focus on: (i) improving the accuracy and granularity of existing data, such as vehicle mileage and fuel consumption, (ii) collecting additional data, including vehicle drivers, time, location, and ``engine hours'' (the total time a vehicle is in use). Available metrics such as age and mileage are imperfect measures of usage of many vehicles, such as police vehicles which may simply idle for long periods of time during police shifts in cold weather.

\vspace{0.1cm}
\noindent \textbf{Challenges.} This collaboration demonstrates a small sample of the insights that can be gained from detailed multivariate analysis of municipal data, but it also illustrates several of the challenges of working with such data. Many aspects of the data---its observational nature; overlapping or difficult-to-decipher descriptions; error and incompleteness which are likely systematic and non-random\footnote{For example, technicians subjectively choose between several job codes: i.e., ``Adjust brakes'' vs. ``Repair brakes'' vs. ``Overhaul brakes''; many older vehicles and jobs are believed to be missing from this data.}---underscore the challenges of working with real-world municipal data often generated as ``data exhaust'' and not with the express aim of providing insights or accurate measurements. Additionally, the distance between our analytical team and the users generating the data (vehicle drivers, technicians, and clerical staff) highlights how challenging it can be to understand data context.

\vspace{0.1cm}
Despite the challenges, even basic insights garnered from a similar analysis can yield significant improvements the status quo for budget-strained municipalities with limited data analysis resources, such as Detroit, and the methods presented here have the potential to apply to a much wider variety of applied data science problems regarding municipal or vehicle fleet data. This work will serve as a model for future municipal-academic research partnerships.  

\section*{Acknowledgements}
{\small
This work is partially supported by National Science Foundation, grant IIS 1845491, and Army Young Investigator Award No. W911NF1810397.
The authors recognize the support of Michigan Institute for Data Science (MIDAS). We would like to thank the General Services Department of the City of Detroit for bringing this project to our attention and making the data available for use.
}

\balance
\bibliographystyle{IEEEtran}
\bibliography{D4GX}

\clearpage 
\appendices

\section{Online Supplementary Results}

The full set of results for the PARAFAC analysis applied to our dataset, consisting of all $R=25$ three-way plots for both the \textit{absolute-time} and the \textit{vehicle-lifetime} analysis, are available in the git repository published with this work. 

\begin{itemize}
    \item \textbf{\textit{Absolute-Time Analysis:}} \url{https://github.com/jpgard/driving-with-data-detroit/tree/master/img/3_way_plots/month_year_log} 
    \item \textbf{\textit{Vehicle-Lifetime Analysis:}} \url{https://github.com/jpgard/driving-with-data-detroit/blob/master/img/3_way_plots/vehicle_year_log/README.md}
\end{itemize}

\section{Algorithms}

\subsection{Bayesian Gaussian Mixture Model (BGMM)}\label{sec:appx-bgmm}

For estimating the in-group for each component of each factor using the loading vectors $\mathbf{a}_r, \mathbf{b}_r, \mathbf{c}_r$, we use a two-component Bayesian Gaussian Mixture Model (BGMM). For each PARAFAC factor $r$, the BGMM is fit directly to the single-valued vectors $\mathbf{a}_r, \mathbf{b}_r, \mathbf{c}_r$. the BGMM is used to assign binary labels to each observation labeling it as either in-group or out-group for a given factor $r$, where the in-group is the cluster with the higher posterior mean. Validation of the model by detailed inspection demonstrated that BGMM achieved the intended result of largely forming clusters of near-zero and non-zero observations. 

 We use a standard finite mixture model from \textit{scikit-learn} with two components and a Dirichlet distribution and a standard weight concentration prior of $\gamma = \frac{1}{2}$, but we note that the model was largely insensitive to the value of $\gamma$ used due to the relatively clean separation of most vectors into zero and non-zero values.

\subsection{Bayesian Difference in Proportions Test (BDPT)}\label{sec:appx-bdpt}

This section describes the Bayesian Difference in Proportions Test (BDPT) in detail. The aim of BDPT is to determine whether there is a true and practically significant difference in the frequency of occurrence of an event between two disjoint populations. The BDPT is implemented with the following hierarchical Bayesian model:
\begin{align}
    \theta_i &\sim \textrm{Beta}(1,1)  \label{eq:theta} \\
    y_i &\sim \textrm{Binomial}(n_i,\theta_i)
\end{align} 
where $i$ denotes two groups of interest (InGroup or OutGroup), $n_i$ indicates the number of observations in each group, and $\theta_i$ indicates the Beta variable drawn in \eqref{eq:theta}. This model is used to estimate both the \textit{difference} in the probability of occurrence between the two groups, $\theta_{\textrm{InGroup}} - \theta_{\textrm{OutGroup}}$, and also the probability that this difference is larger than a prespecified Region of Practical Equivalence, or ROPE \cite{Kruschke2011-cz}, which is equivalent to estimating 

\begin{equation}
    \mathbb{P} \Big( \theta_{\textrm{InGroup}} - \theta_{\textrm{OutGroup}} \notin \textrm{ROPE} \Big).
\end{equation}

We implement this test using the Python package \texttt{pymc3}, using two chains of 2000 MCMC samples each with a burn-in period to perform posterior inference. This relatively small sampling was determined to be acceptable given the simple model, which achieved good MCMC convergence.

\section{Model Implementation and Hyperparameters}

\subsection{PARAFAC}\label{sec:appx-parafac}

We use the PARAFAC implementation in the MATLAB Tensor Toolbox. Specifically, we utilize the \texttt{cp\_nmu()} function to compute the PARAFAC decomposition, which implements the NMF algorithm of \cite{lee2001algorithms}. This algorithm uses a mutiplicative update to minimize the reconstruction error between a data matrix $X$ and its reconstruction $P$ by minimizing the square of the Euclidean distance between $X$ and $P$, solving the problem
\begin{align}
    \textrm{min}_{P} ~ ||X - P||^2 = \textrm{min}_{P} ~  \sum_{ij}(X_{ij} - P_{ij})^2
\end{align}
where $P$ is a nonnegative factorization of the matrix $X$.

For our experiments, we use a tolerance of $10^{-4}$ and a maximum of 500 iterations; however, with $R=25$ factors, the tolerance is reached in far fewer than the maximum number of allowed iterations.

\begin{figure}
    \centering
    \includegraphics[width = 0.8 \columnwidth]{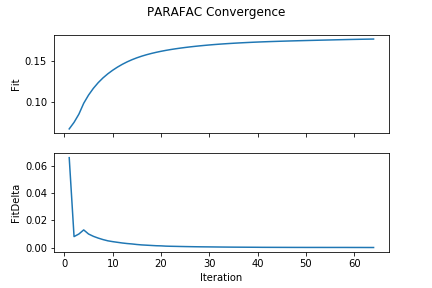}
    \caption{Top: PARAFAC goodness-of-fit metric \eqref{eqn:goodness-of-fit} over training iterations. Bottom: Convergence measured by change in \eqref{eqn:goodness-of-fit} over iterations.}
    \label{fig:parafac-convergence}
\end{figure}

Figure \ref{fig:parafac-convergence} shows convergence and fit diagnostics for the PARAFAC model. The upper panel shows a goodness-of-fit metric, 
\begin{align}
    1 - \frac{\sqrt{\sigma_{max}(X) + \sigma_{max}(P) - 2 \cdot \langle X, P \rangle }}{ ||X||}\label{eqn:goodness-of-fit}
\end{align}
where $\sigma_{max}(\cdot)$ indicates the largest singular value of a matrix, and $X$ and $P$ indicate the data matrix and the PARAFAC reconstruction, respectively, as computed by \cite{Bader_undated-ue}. Note that the maximum possible value of this metric is 1, indicating a perfect reconstruction, although what qualifies as an acceptable value of this metric is application-dependent.

The lower panel of Figure \ref{fig:parafac-convergence} shows the change in \eqref{eqn:goodness-of-fit} over iterations. While the PARAFAC algorithm is only guaranteed to converge to a local minima \cite{lee2001algorithms} and global optimality cannot be guaranteed, our results indicate smooth and stable convergence.

\subsection{LSTM Sequence Prediction Model}\label{sec:appx-lstm}

Our LSTM model is a 2-layer LSTM which considers up to 20 previous items in the sequence, if they exist, when predicting the next job. This model uses a 200-dimensional dense representation of the input features, which allows it to learn about relationships between repairs to different systems. 

The model uses the following hyperparameters:

\begin{itemize}
    \item Gradient descent optimizer; initial learning rate $= 1.0$.
    \item Learning rate decay by factor of 0.5 after completion of the first 4 epochs.
    \item Context window size = 20
    \item Hidden unit size $=200$
    \item Batch size $=20$
    \item Max gradient norm $=5.0$
\end{itemize}

The model is implemented in Tensorflow 1.x. Training on our dataset completes in less than 10 minutes on a standard laptop CPU.

\subsection{ARIMA Cost Forecasting Model}\label{sec:appx-arima}

ARIMA has three free parameters, all of which are intuitive to set: $p$, $d$, and $q$, indicating the number of autoregressive terms, the degree of differencing to remove trends from the time series, and the order of the moving average, respectively. 

Our model uses $p = 6$ autoregressive terms, meaning that it explains each month's average cost based on values from the previous 6 months; and $q = 4$ moving average terms. $p$ and $q$ are tuned to minimize the AIC and BIC scores when fitted to the data. We use $d = 2$ as the degree of differencing and do not tune this parameter, as second-order differencing is standard for removing trends and seasonality from time series data.

The models are implemented in R, version 3.x, using the \texttt{arima} and \texttt{auto.arima} functions.

\section{Open-Source Implementation of Differential Sequence Mining}

As a part of the contribution of this paper, we have made available an open-source implementation of the PRISM algorithm used. This includes the Bayesian Difference in Proportions Test (BDPT), as well as our implementation of the original frequentist differential sequence mining method used in \cite{Kinnebrew2013-ny} and the relevant utility functions. 

Due to a non-disclosure agreement with the City of Detroit, the data itself cannot be made publicly available. A stable implementation of PRISM in Python, including Python, MATLAB, and R code to reproduce the full analysis on a new dataset, is available at \url{https://github.com/jpgard/driving-with-data-detroit}. Installation instructions are available in the repository.

\end{document}